# Experimental studies of instabilities of laminar premixed flames

Geoff Searby

CNRS / Université de Provence

Institut de Recherche sur les Phénomènes Hors Équilibre

49 rue Frédéric Joliot-Curie

13384 MARSEILLES Cedex 13

France

e-mail : Geoff.Searby@irphe.univ-mrs.fr

**Abstract**    We first briefly recall the basic mechanisms controlling the hydrodynamic and thermo-diffusive stability of planar laminar premixed flames, and give the state of the theoretical analysis. We then describe some novel experiments to observe and measure the growth rate of cellular structures on initially planar flames. The first experiment concerns the observation of the temporal growth of wrinkling on a freely propagating planar flame. A second experiment concerns the spatio-temporal growth of structures of controlled wavelength on an anchored flame. The experimental observations are compared to theoretical dispersion relation. Finally, we compare observations of the non-linear evolution to saturation with the predictions of an extended Michelson-Sivashinsky equation.

## Introduction

In general, planar flame sheets are unstable to perturbations and will spontaneously form cellular structures. Darrieus[1] in 1938 was the first to recognize that the gas expansion produced by heat release in a wrinkled premixed flame will deviate the flow lines in the front

---

towards the normal to the flame. Since the Mach number, $M \equiv S_L/c$, (where $S_L$ is the laminar flame speed and $c$ is the speed of sound) of deflagrations is very small, the flow is quasi-incompressible. Hence the upstream flow lines are also deviated, creating flow divergence and velocity gradients that will increase the wrinkling of the flame (Fig. 1).

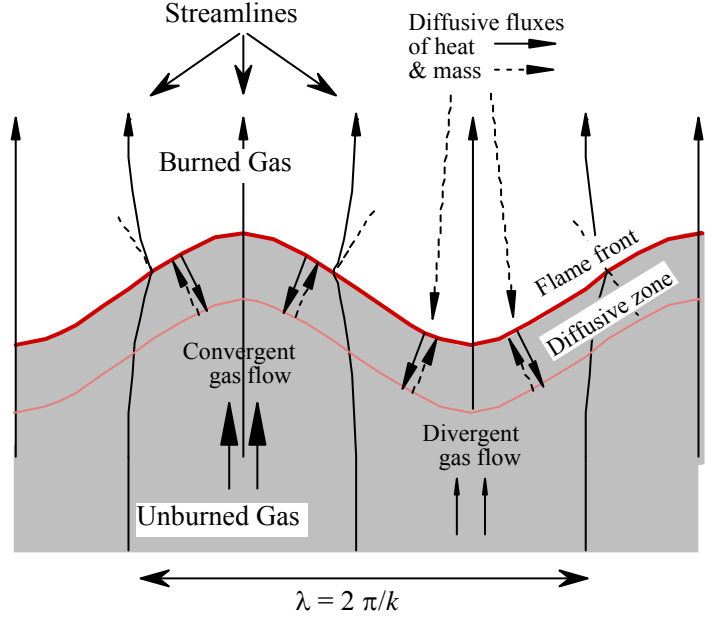

Figure 1. Structure of a wrinkled flame front showing the hydrodynamic streamlines and the diffusive fluxes of heat and mass.

This unconditional hydrodynamic instability was predicted independently by Landau [2] in 1944. Considering the flame as a thin interface between unburned and burned gases, he found that the growth rate, $\sigma$, of this hydrodynamic instability should vary as

$$\sigma = k \cdot S_L \cdot \left( \frac{E}{E+1} \sqrt{\frac{E^2 + E - 1}{E}} - 1 \right), \qquad (1)$$

where $k$ is the wavenumber of the perturbation and $E = \rho_u/\rho_b$ is the gas expansion ratio. A more complete description can be found in the work of Zeldovich et al.[3].

If the finite thickness of the flame front is considered, $\delta = D_{th}/S_L$, where $D_{th}$ is the thermal diffusivity, then the effect of wrinkling will induce transverse heat and mass fluxes that alter the local flame temperature and lead to modifications of the local flame speed [3-5]. A purely one-dimensional diffusive instability of premixed laminar flames (pulsating propagation), has been predicted [6, 7] for mixtures having a very high Lewis number coupled with a very high activation energy. However the theoretical threshold value of these parameters is beyond the range of real flames and this type of instability has never been observed experimentally.

The combined effects of preferential diffusion and hydrodynamic instability has been studied some twenty years ago by many authors [8-11]. The approximation of high activation energy and multi-scale asymptotics were used to solve the problem at three physically distinct scales, which are the inner scale of the chemical reaction zone (scale $\delta/\beta$), the scale of the diffusion zone (scale $\delta$) and the outer scale of the hydrodynamic zone (scale $k^{-1}$). Here, $\beta \approx 10$ is the reduced activation energy, or Zel'dovich number:

$$\beta = \frac{E}{kT_b} \frac{(T_u - T_u)}{T_b} \qquad (2)$$

The effect of gravity acting on two fluids of different density, burned and unburned gas, was also included on the analysis. The effect of gravity can help stabilise flames when the light burned gas is above the heavy unburned gas. The most complete description of laminar flame stability is given in a little known 1983 paper of Clavin and Garcia [12]. These authors solve the dynamics of small amplitude wrinkling of premixed flames including the effect of temperature dependent diffusion coefficients. They find that the dispersion relation for the growth rate, $\sigma$, of small amplitude wrinkling with wavenumber $k$ is given by:

$$(\sigma \tau_t)^2 A(k) + \sigma \tau_t B(k) + C(k) = 0 \qquad (3)$$

The growth rate of the instability is given by the real part of $\sigma$. The wavenumber dependent coefficients in (3) are given by :

$$A(k) = \frac{E+1}{E} + \frac{E-1}{E} k \delta \left( Ma - J \frac{E}{E-1} \right),$$

$$B(k) = 2 k \delta + 2 E (k\delta)^2 (Ma - J),$$

$$C(k) = \frac{E-1}{E} \frac{k \delta}{Fr} - (E-1)(k \delta)^2 \left[ 1 + \frac{1}{E \, Fr} \left( Ma - J \frac{E}{E-1} \right) \right]$$

$$+ (E-1)(k \delta)^3 \left[ h_b + \frac{3E-1}{E-1} Ma - \frac{E}{E-1} 2J + (2\,Pr - 1) H \right]$$

(4)

Here, $\delta = D_{th} / S_L$ is the thickness of the thermal diffusion zone, $\tau_t = \delta / S_L$ is the flame transit time, $Fr = S_L^2 /(g\,\delta)$ is the Froude number of the flame, $g$ is the acceleration of gravity (positive when the flame propagates downwards) and $Pr = \nu / D_{th}$ is the Prandtl number. The quantities $J$ and $H$ account for the temperature dependence of the diffusivities. They are given by the integrals:

$$H = \int_0^1 (h_b - h(\theta))\,d\theta,$$

$$J = (E-1) \int_0^1 \frac{h(\theta)}{1 + (E-1)\,\theta}\,d\theta$$

(5)

where $\theta = (T - T_u)/(T_b - T_u)$ is the reduced temperature, $h(\theta)$ is the thermal diffusivity times density ($\rho D_{th}$) at temperature $\theta$, normalised by its value in the unburned gas, and $h_b$ is the value of $h(\theta)$ in the burned gas. The above relations were obtained in the linearised limit of long wavelength low frequency perturbations $k \delta \ll 1$, $\sigma \tau_t \ll 1$ with small amplitude of wrinkling. The expressions are accurate to order $(k \delta)^2$ and $(\sigma \tau_t)^2$.

The sensitivity of the local burning velocity to curvature and stretch is contained in the Markstein number, $Ma$, [8]. When the Markstein number is positive, which is generally the case, the effect of curvature is to decrease the burning velocity of a region that is convex towards the unburned gas, implying that the flame is thermo-diffusively stable. Clavin and Garcia [12] have given an expression for the Markstein number for the simplified case of an overall one-step chemical reaction controlled by an Arrhenius law:

$$Ma = \frac{E}{E-1} J + \frac{\beta}{2} \int_0^1 \frac{h(\theta) \ln(\theta^{-1})}{1+(E-1)\theta} d\theta \qquad (6)$$

In (6), $\beta$ is the reduced activation energy, or Zel'dovich number. The first term on the r.h.s. of (6) arises from transverse *convective* transport of heat and species within the diffusive thickness of the wrinkled flame. The second term arises from transverse *diffusive* transport of heat and mass. Equation (6) shows that when convection within the flame thickness is also considered, a flame can be thermo-diffusively stable ($Ma > 1$) even for Lewis numbers smaller than unity.

The approximation of one-step chemistry is not good and in general and (6) should be considered only as a rough approximation. It is generally necessary to obtain Markstein numbers from experimental measurements [13-16] or from direct numerical simulations [17, 18].

Figure 2 shows typical plots of the real part of the reduced growth rate, $\sigma \tau_t$, calculated as a function of the reduced wavenumber, $k\delta$, for propane air flames. For small wavenumbers, the growth rate first increases linearly with the wavenumber, as predicted by Darrieus and Landau. Then, at larger wavenumbers, the growth rate decreases with the square of the wavenumber, due to thermo-diffusive effects. The effect of gravity is to shift the growth rate curve downwards and to reduce the range of instable wavenumbers. The effect is only significant for slow flames, $S_L < 0.2$ m/s. For sufficient slow flames, $S_L < 0.12$ m/s, the effect of gravity can stabilise downwards propagating flames at all wavenumbers [11, 13].

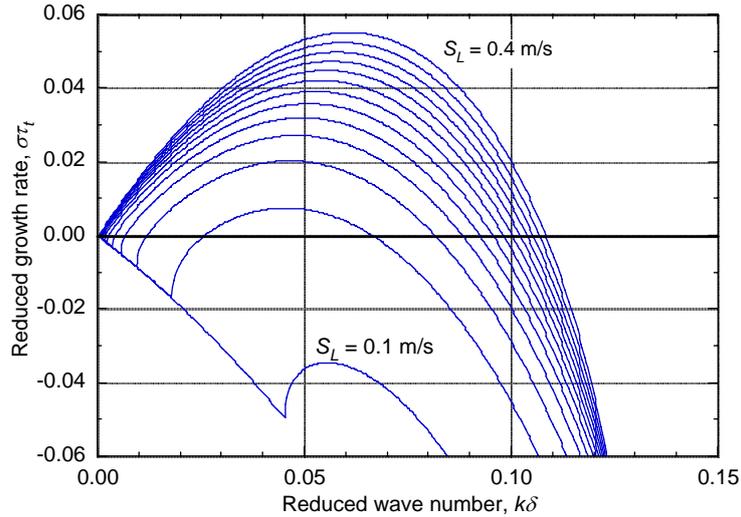

Figure 2. Reduced growth rate of Darrieus-Landau instability plotted as a function of reduced wavenumber for flame speeds in the range $0.1 \leq S_L \leq 0.4$ m/s. (lower to upper in steps of 0.025 m/s). Other parameters appropriate for lean propane flames. The data needed in equations (3-5) were taken from the CHEMKIN package [19]. *Ma* = 4.5 [13].

This linear analysis is expected to be correct during the linear part of the growth of the Darrieus-Landau instability. However, until recently, there has been no direct experimental verification of the theory. The reason for this lies in the difficulty of producing the initial condition of a planar, freely propagating, unstable flame. The characteristic growth time of the instability is typically 20-50 ms, which is short compared to the time needed to establish a freely propagating flame of finite dimensions. It follows that the Darrieus-Landau instability has nearly always been observed in the phase of non-linear saturation.

**Experimental Verification on Planar Flames**

The first direct experimental verification of (2) was performed by Clanet and Searby [20]. They used a novel technique of acoustic restabilisation to produce a perfectly planar laminar flame that is otherwise intrinsically unstable. The imposed acoustic field was then removed on a short time scale, (about 1 ms), and the unconstrained growth of the Darrieus-Landau instability was observed. The contents of this section will describe their work.

**Effect of Acoustic Field on a Laminar Flame**

The effect of an imposed acoustic field on a freely propagating premixed laminar flame front has been given by Searbyand Rochwerger [21]. In the approximation $a\,k \ll 1$, where $a$ is the amplitude of the wrinkling, these authors have shown that the dynamics of a flame front in an imposed acoustic field characterised by the frequency, $\omega_a$, and displacement velocity, $u_a$, is dominated by the effect of the periodic acoustic acceleration acting on the flame seen as a thin interface between two fluids of different density. The dynamics of this system can be obtained from (3) and (4) by introducing this periodic acoustic acceleration into the Froude number, along with the acceleration of gravity:

$$Fr^{-1} = \frac{g\,\delta}{S_L^2} - \omega_a\,u_a \cos(\omega_a\,t) \qquad (7)$$

Writing $\alpha(y,t) = a\exp(\sigma t - i\,k\,y)$, where $\alpha$ is the amplitude of wrinkling at wavenumber $k$, it can be seen that (3) is the time Fourier transform of the following equation of motion for $\alpha$:

$$A(k)\frac{\partial^2 \alpha}{\partial t^2} + B(k)\frac{\partial \alpha}{\partial t} + \left[C(k) - C_1(k)\cos(\omega_a t)\right]\alpha = 0 \qquad (8)$$

Where the wavenumber dependent coefficients $A(k)$, $B(k)$ and $C(k)$ are given in (4). The acoustic acceleration has introduced an extra term $C_1(k)$:

$$C_1(k) = \frac{E-1}{E}k\,\delta\,\omega_a\,u_a\left[1 - k\,\delta\left(Ma - J\frac{E}{E-1}\right)\right] \qquad (9)$$

Equation (8) is formally equivalent to a parametrically driven damped harmonic oscillator in which the acoustic acceleration appears as the driving force. This equation, (8), can be reduced to the Mathieu equation [22] by a simple substitution of variables [8]. The solutions of this latter are known to present regions of instability separated by a stable domain, Details are given in [21].

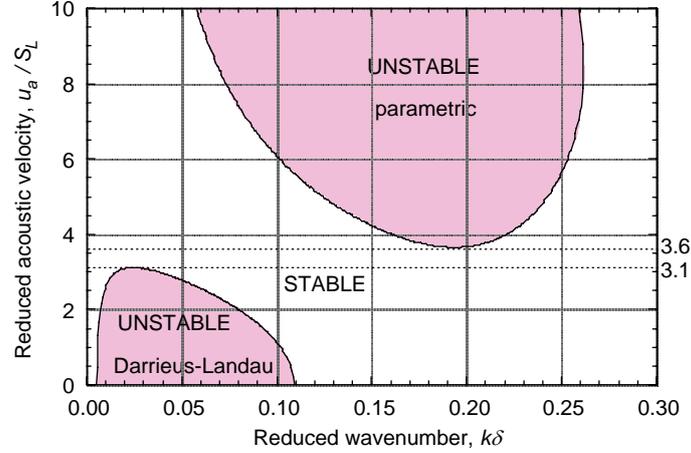

Figure 3. Regions of stability and instability of lean premixed propane flame subjected to a periodic acoustic acceleration.

It may be noted that the flame front does not have a single characteristic frequency, but a continuum associated with the continuum of wavelengths that can be excited on its surface. In Fig. (3), we have plotted a typical stability diagram for the solutions to equ (8), using parameters appropriate for a lean propane-air flame. The stability thresholds are plotted as a function of reduced wavenumber for the reduced frequency $\omega \tau_t = 0.88$. The lower pocket of instability, at small wavenumber and low acoustic amplitudes corresponds to the Darrieus-Landau instability including thermo-diffusive effects. A remarkable feature is that the limits of this unstable domain moves towards each other as the acoustic amplitude is increased from zero. For an acoustic amplitude $u_a / S_L \approx 3$, the upper and lower wavelength limits of this unstable zone merge, and the planar flame front is *restabilised* at all wavenumbers with respect to the hydrodynamic instability. At higher acoustic levels, there is a second domain of instability. In this domain, the flame structures oscillate at half the acoustic frequency; it is the domain of *parametric* instability. The fact that a finite value of excitation is needed to excite this instability is related to the presence of a damping term in

equ.(8). The lower threshold of this unstable domain is a function of the reduced acoustic frequency. As the reduced frequency is decreased (or equivalently, as the laminar flame speed is increased), the lower threshold moves downwards and will overlap the lower instability domain. However, for excitation frequencies of a few hundred Hertz and for moderate flames speeds less than 0.2 m/s, the flame can be stable at all wavelengths for a small range of acoustic excitations having acoustic displacement velocities $u_a \approx 3.5\,S_L$. This corresponds to an acoustic intensity of about 140 dB. See reference [21]. This acoustic stabilisation window has been used to maintain an intrinsically unstable flame in a planar state, prior to observing the growth of the Darrieus-Landau instability.

**Experimental Set-up for Acoustically Stabilised Flame**

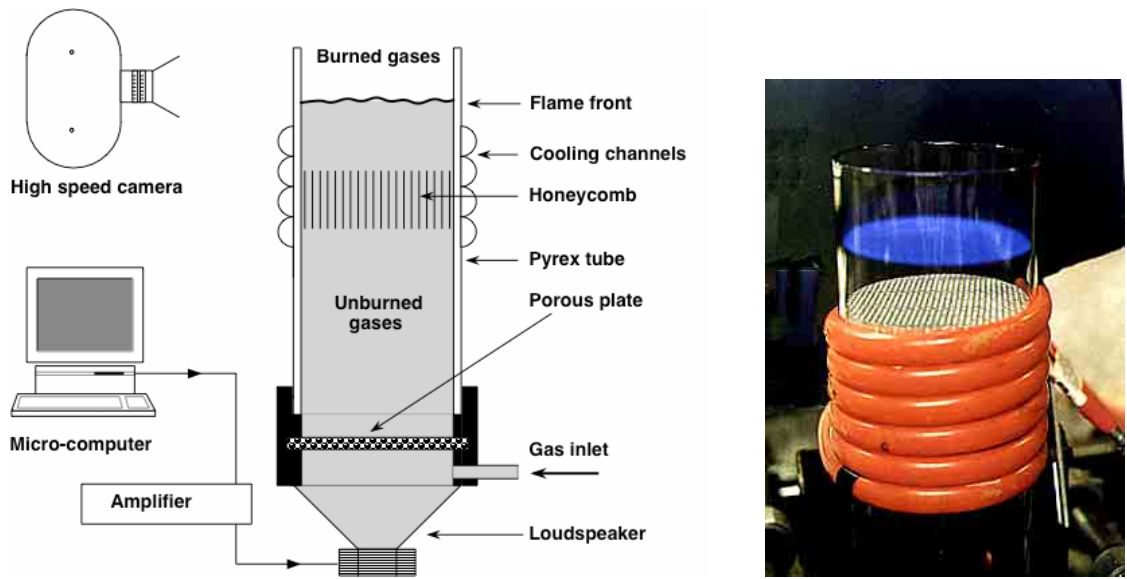

Figure 4. Experimental set up used to prepare an acoustically stabilised planar premixed flame and photograph of an acoustically stabilised flame.

The experimental apparatus is shown in Fig. 4. The premixed gases are fed into the bottom of a Pyrex glass tube 100 mm diameter and 400 mm long, just below a 50μm porous plate whose role was to laminarise the flow. The flame was held stationary in the laboratory frame, about 50 mm below the tube exit, by careful adjustment of the gas flow rate. A 2 mm aluminium honeycomb structure 40 mm long was placed a few centimetres upstream from the flame to help maintain a laminar and homogeneous gas flow. A helical cooling tube was wound round the outside of the Pyrex tube, below the flame front, in order to prevent wall heating by heat conduction and radiation from the flame. The Pyrex tube was closed at its lower extremity by a loudspeaker, which was fed from a micro-computer used as a programmable signal generator. The acoustic impedance of the porous plate is high and the Pyrex tube behaved as an open-closed resonator. It was excited in the 1/4 wavelength longitudinal mode (230 Hz). The natural acoustic damping time (≈12 ms) was not always short compared to the growth time of the Darrieus-Landau instability. To overcome this limitation, at the start of a measurement, the damping of the acoustic standing wave was increased artificially by injecting one cycle of a signal in phase opposition with the pressure in the tube. An intensified high-speed cine camera recorded the luminous emission from the flame.

Experiments were performed with lean propane-air mixtures with equivalence ratios in the range $0.56 \leq \Phi \leq 0.67$, corresponding to burning velocities in the range 0.11 m/s $\leq S_L \leq$ 0.20 m/s. For leaner flames, the front was intrinsically stable at all wavelengths. For faster flames, a flat laminar flame could not be obtained by this method. In order to control the wavelength and orientation of the cellular structures that developed when the acoustic stabilisation was removed, the upstream gas flow was slightly perturbed by placing an array of parallel wires, 2 mm in diameter, on the downstream face of the honeycomb. The object of this scheme was to excite purely 2-D cells at a chosen wavelength of 2 cm, close to the most unstable wavelength, see Fig.2. The luminous emission from the flame front was filmed edge-on in a direction parallel to the axis of the cells. Fig.5 shows a sequence of typical images taken from a high speed film during the growth of the Darrieus-Landau instability. The time $t = 0$ corresponds to the instant at which the acoustic field was removed.

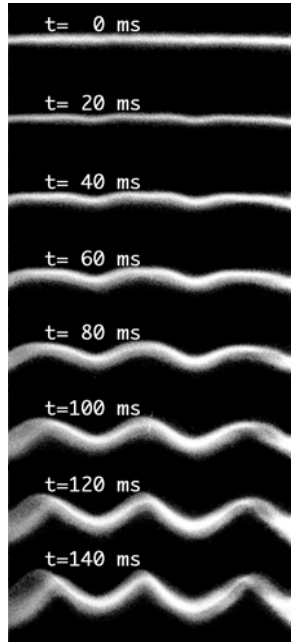

Figure 5. Images taken from a high speed film of the growth of instability. Framing rate 500 i/s. Wavelength = 2 cm. Flame speed = 0.115 m/s.

The apparent thickening of the flame, particularly at high cell aspect ratios indicates the presence of slight three-dimensionality of the wrinkling. The peak to peak amplitude of the wrinkling was measured on digitized images and plotted in semi-log coordinates as shown in Fig.6.

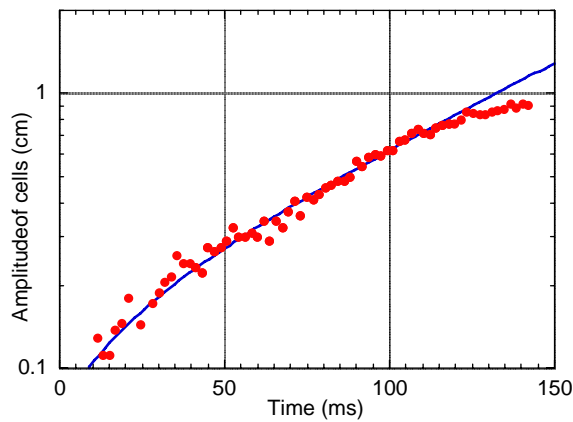

Figure 6. Semi-log plot of the amplitude of the cellular structures of Fig.5. Fitted curve from equ.10.

The large scatter of the points in the early stages of the growth arise from the small amplitude of the cells, of the order of the apparent flame thickness. The non-linearity at long times indicates the onset of saturation of the instability. This nonlinearity of the shape of the cells is clearly visible in Fig.5 after 140 ms. These points were systematically eliminated before data reduction to obtain the growth rate. The points were fitted to an exponential function of the form:

$$a(t) = \frac{1}{2}\left[\left(a_0 + \frac{v_0}{\sigma}\right)\exp(+\sigma t) + \left(a_0 - \frac{v_0}{\sigma}\right)\exp(-\sigma t)\right], \qquad (10)$$

which is the general solution of $\partial^2 a / \partial t^2 = \sigma^2 a$ with the initial conditions $a(0) = a_0$ and $\partial a(0)/\partial t = v_0$. Here, $v_0$ is the rate of increase of the wrinkling at time $t = 0$, supposed equal to the measured peak-to-peak velocity modulation produced by the wires in the flow. The precision of the growth rate obtained by this procedure was $\pm 5$ s$^{-1}$. These experimentally measured growth rates are plotted as a function of laminar flame speed in Fig.7.

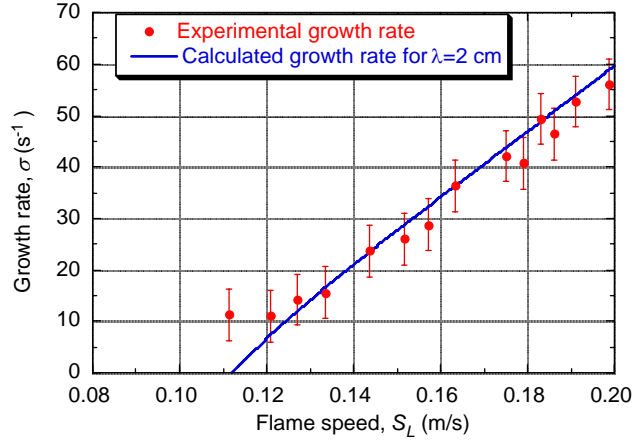

Figure 7. Comparison of experimental and calculated growth rates. The flame speeds, $S_L$, were taken from [23] and gas expansion ratios were calculated using the CHEMKIN package [19].

All measurements were made at a fixed (forced) wavelength of 2 cm. The full line in Fig.7 shows the calculated growth rates obtained from the dispersion relation (3). A Markstein number of 4 was found to give best agreement with the experimental data. This value of 4, for lean propane-air mixtures may be compared with the value 5 found experimentally by Tseng *et al.* [14], the value of  and the value 4.3 found previously by Searby and Quinard [13].

**Experimental Verification of Growth Rate on Inclined Flames**

In the previous section we have described an experimental technique that has been used to observe the temporal growth rate of the Darrieus-Landau instability on an initially planar flame. Because of limitations inherent to the technique, the observations were limited to relatively slow flames, $S_L$ < 0.2 mm/s and to observation at a wavelength close to that of maximum growth rate. In this section we will describe a different experimental technique used by Truffaut and Searby [24, 25] to measure the growth rate of the instability on an inclined flame over a wider range of wavenumbers and flame speeds.

In these experiments a small amplitude oscillation, at frequency $f$, was imposed at base of the inclined flame and the resulting structure was convected downstream by the tangential component of the gas flow. The wavelength of the structure is $\lambda = u_{//} / f$ where $u_{//}$ is the tangential convection velocity. In this configuration the growth of the instability is spatio-temporal and not simply temporal. The spatial growth rates were converted to temporal growth rates using a Lagrangian time obtained from the displacement velocity of the wrinkles, $u_{//}$. This conversion is expected to be valid if adjacent cells have (nearly) the same size. This condition is satisfied when the growth time is long compared to the time taken to convect the pattern a distance of one wavelength downstream, i.e. for $\Re[\sigma^{-1}] \gg f^{-1}$. Thus (3)-(5) were again used to analyse the results of the experimental measurements.

Equations (3)-(5) are strictly valid only for planar flames propagating upwards or downwards. The case of an inclined flame in a gravity field has been treated theoretically by Garcia and Borghi [26]. The effect of gravity is to introduce a bulk force with a component parallel to the flame front and the problem becomes asymmetric. In a simplified analysis,

Garcia and Borghi found that 2-D wrinkles, whose axes of wrinkling are horizontal, can have a finite propagation velocity, different from the convective velocity. According to their analysis, this propagation velocity goes to zero for horizontal and for vertical flame fronts. In the experiments of Searby and Truffaut, the flame fronts are close to vertical (< 5°) and moreover the Froude number of the flames, $Fr = S_L^2/(g\delta)$ was always very large ($Fr > 90$). The effect of gravity is then expected to be negligible. Experimentally, it was found that the measured displacement velocity of the wrinkles was always equal to the velocity of the tangential gas flow.

**Inverted-'V' Burner and Flame Excitation Device**

A laminar slot burner was used to produce a two-dimensional inverted-'V' premixed flame and an electrostatic deflection system was used to impose the wavelength of the perturbation.

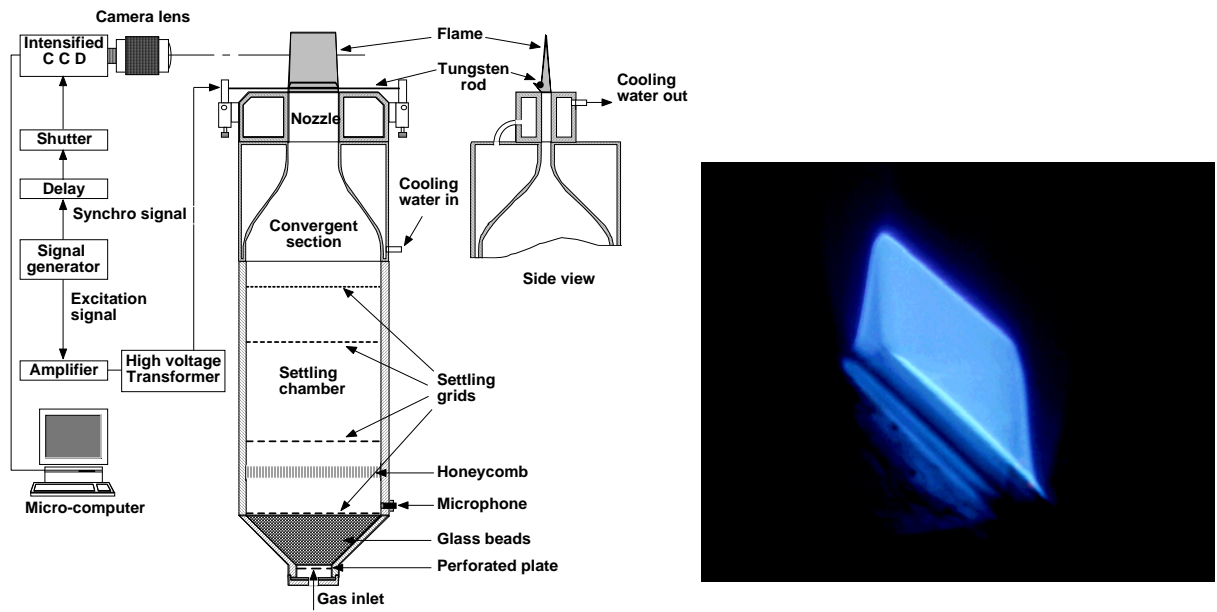

Figure 8. Schematic diagram of the inverted-'V' burner and excitation system.
Right, photograph of 2-D flame

The experimental apparatus is shown schematically in Fig.8. The premixed propane-air-oxygen gas was fed to the bottom of a burner designed to produce a laminar "top-hat" velocity profile at the exit. The burner comprised a divergent section, a settling chamber, a convergent section and a nozzle. The 60 mm high divergent section was filled with 4 mm diameter glass beads to break up the incoming flow. The flow was then laminarised in a 140 mm square settling chamber containing an aluminium honeycomb, followed by three metal grids of decreasing mesh size. Finally a 2-D convergent section, with a contraction ratio of 30:1, accelerated the flow up to the 8x80 mm exit section and reduced the residual velocity fluctuations to a small fraction of the mean flow velocity. The 1:10 aspect ratio of the exit provided a 2-D inverted-'V' flame. End effects perturbed the flame for less than 10 mm at each extremity. The thickness of the viscous boundary layers was less than 1.5 mm and the transverse velocity profiles were flat to better than 1 % over the central region of 5 mm. The residual turbulence was less than 0.5 %. In order to minimise the development of a thermal boundary layer near the walls, both the convergent section and nozzle were water-cooled at the temperature of the unburnt mixture (20°C).

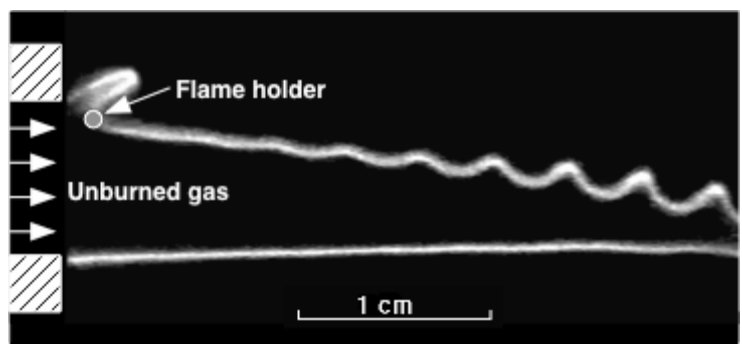

Figure 9. Instantaneous image of growth of Darrieus-Landau instability of a propane-air-oxygen flame, equivalence ratio = 1.33, 28 % oxygen, flow velocity 8.56 m/s, excitation frequency 2100 Hz.

One side of the flame was anchored on a thin tungsten rod, 0.6 mm diameter, placed just above and parallel to the burner exit, see Fig.9. The laminar flame was perturbed on one face by applying an alternating high voltage between the rod and the burner exit. The resulting electric field displaces the flame locally and produced a periodic sinusoidal 2-D wrinkle on the flame front as shown in Figs.8 and 9. The axis of the wrinkle was parallel to the burner slot. The wrinkle was convected downstream by the gas flow and its amplitude was observed to increase exponentially. This system permitted a precise control of both the initial amplitude and of the wavelength of the wrinkle through control of the amplitude and of the frequency of the applied signal. Typical ranges of voltage and frequency were 2-4 kV and 1-4 kHz respectively.

A similar electrostatic technique was first used by Polanyi and Markstein in 1947 [27]. However the present system is slightly different, in that Markstein placed a deflection electrode in the burned gas, whereas we used the electrode as a flame holder. We have observed that, for a given voltage, the amplitude of small wavelength wrinkles was much greater when the flame was excited via a flame-holder, probably because the electric field acts more locally on the flame front. The deflection of a flame in an electric field is attributed to a body force produced by momentum transfer between charged particles and neutral molecules. Some authors [28, 29] have argued that the momentum transfer from heavy ions dominates momentum transfer from electrons, giving rise to a non-zero body force in the direction of the electric field.

Experiments were performed on rich propane-air flames and propane-air-oxygen flames with equivalence ratios in the range $1.05 \leq \Phi \leq 1.33$. The corresponding laminar flame velocities were in the range 0.43 m/s $\geq S_L \geq$ 0.27 m/s [23] for the propane-air flames. The velocities were 0.69 m/s and 0.51 m/s for flames with 28 % oxygen at equivalence ratios of 1.05 and 1.33. The gas flow velocities were 7.4 m/s, 6.05 m/s and 8.56 m/s.

**Data Acquisition**

The wrinkled flame front was observed using a short exposure intensified CCD camera, viewing parallel to the long edge of the burner slot. In this configuration the flame is a long 3–D object. Since we were interested in imaging a cross section of the flame, we used an optical system having a short depth of focus and a high magnification ratio. The camera was focussed in the centre of the burner slot and the depth of focus was roughly 2 mm. This value also represented the uncertainty in the knowledge of the position of the object plane. The resulting relative uncertainty in the magnification of the image is less than 1 %. An inconvenient of this optical system is some unsharpness of the image due to the superposition of the out-of-focus contours from other distances. However the resolution of the images obtained by this method was acceptable, see Fig.9.

The video camera was triggered at a frequency close to 50 Hz by a pulse generator synchronized to an appropriate sub-harmonic of the excitation signal. An adjustable delay allowed us to take images at different phases of the excitation. For all images, the gate-time of the intensifier was less than 100 µs and the optical gain was adjusted manually so as to obtain non-saturated images. The images were digitized at a resolution of 760*570 pixels. The spatial calibration, obtained by imaging a grid placed in the object plane, was 52.6 µm per pixel.

In the experiment described here, the flame front was not freely propagating, but attached to a flame holder in the proximity of a companion front, forming an inverted-'V' flame. In this situation, care must be taken to ensure that the observed perturbations were growing freely. Firstly it was ensured that residual and external flow perturbations were so small that only the imposed wavelength appears on the flame front. This was verified by noticing that, in the absence of electric excitation there was no detectable motion of the flame front. The residual perturbations were so small that they did not have time to grow to measurable amplitude before they were convected to the flame tip. Secondly it was ensured that the two flame fronts did not interact. It is known that the hydrodynamic field ahead of a wrinkled premixed flame is modified over a characteristic distance equal to $k^{-1}$ [30]. If the distance between the two fronts is smaller than this value, the instability developing on one side of the flame can be influenced by the presence of the other flame. To avoid this problem the angle between the two sides of the flame was kept small and their spacing, in the field of visualization, was kept large compared to $k^{-1}$. For this reason the measurements were also restricted to wavelengths less than 6 mm. The absence of interaction between the two flame fronts can be checked from images such as Fig.9. The lower side of the flame shows no perturbation from the wrinkled side of the flame, except at the far right side of the image, close to the flame tip, where the amplitude of the wrinkling is already saturated. It may be concluded that the contrary is also true, the wrinkled side of the flame was not affected by the presence of the unperturbed side.

**Data Processing**

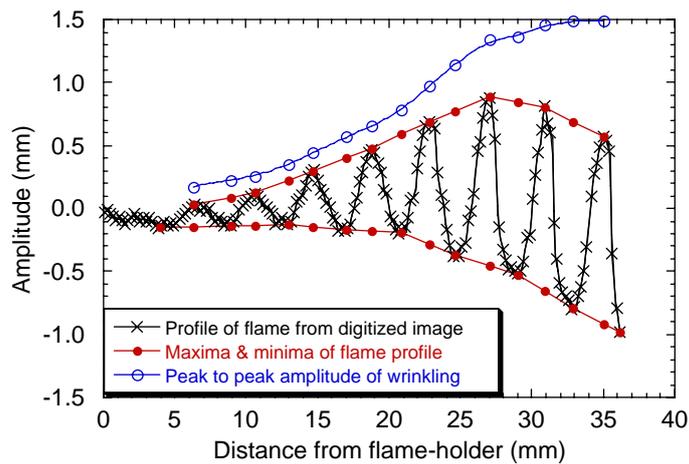

Figure 10. Plot of flame profile from Fig.9 showing method of measuring amplitude of wrinkling and corresponding peak-to-peak amplitude.

Images of the flame, such the one shown in Fig.9, were processed to obtain the amplitude and the wavelength of the wrinkles as a function of the distance downstream. A program searched for the brightest pixel on each vertical line of pixels in a sub-window containing the excited side of the flame. The program then fitted a parabola through the intensity at this point and at two other points on each side of the maximum. The flame position was given by the position of the maximum of the fitted curve. This algorithm gave the position of the flame front as a function of the distance from the flame holder. A typical plot is shown in Fig.10.

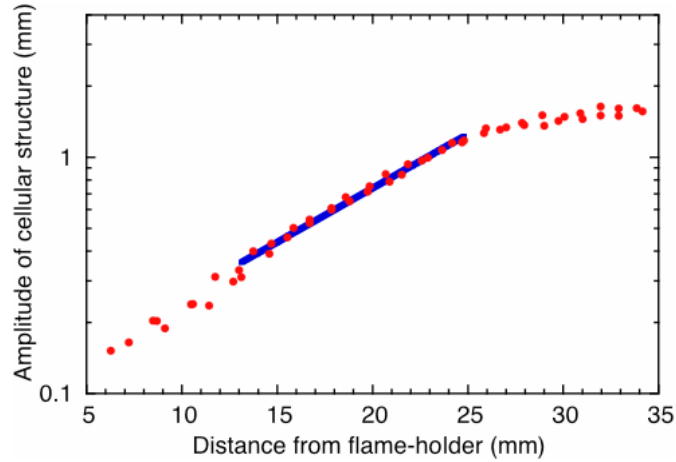

Figure 11. Semi-log plot of peak-to-peak amplitude of flame wrinkling obtained from Fig.10.

After low pass filtering of the curve the maxima and minima were located. The value corresponding to excitation in phase opposition was approximated by interpolation of the preceding and following minima (or maxima), as shown in Fig.10. In this way, the peak-to-peak amplitude and the wavelength of the wrinkle were obtained for each extremum position. For each frequency of excitation this operation was repeated for 4 different phases of the signal. The peak-to-peak amplitude was then plotted in semi-log coordinates, as shown in Fig.11. The spatial growth rate of the instability, $\sigma$, was obtained by fitting this curve with an exponential function. The initial points, close to the flame holder, were influenced by the presence of the rod and also by the presence of the electric field. The amplitude of the wrinkle in this region was of the order of the apparent flame thickness $\delta \approx 0.2$ mm. These points were systematically ignored, along with the downstream points in the region of saturation where the wrinkles are cusped. The amplitude of the wrinkling was measured with a precision of the order of $\delta \approx 0.2$ mm, so there was a resulting uncertainty in the values of the fitted parameters. The spatial growth rate was thus measured with a precision that ranged from 10 % to 30 %, depending on the size of the wrinkles.

**Results**

The experimental results showed that the wrinkles produced at the base of the flame were convected downstream at a constant velocity, $u_{//}$, equal to the tangential flow velocity to within experimental error. The wavelength of excitation is thus given by $\lambda = u_{//} / f$ where $f$ is the excitation frequency. We have not observed any tendency for the wrinkles to have a finite propagation velocity with respect to the flow, as suggested in the work of Garcia and Borghi [26]. The amplitude of the wrinkles was observed to increase exponentially up to the onset of saturation, visible in Figs 9, 10 and 11.

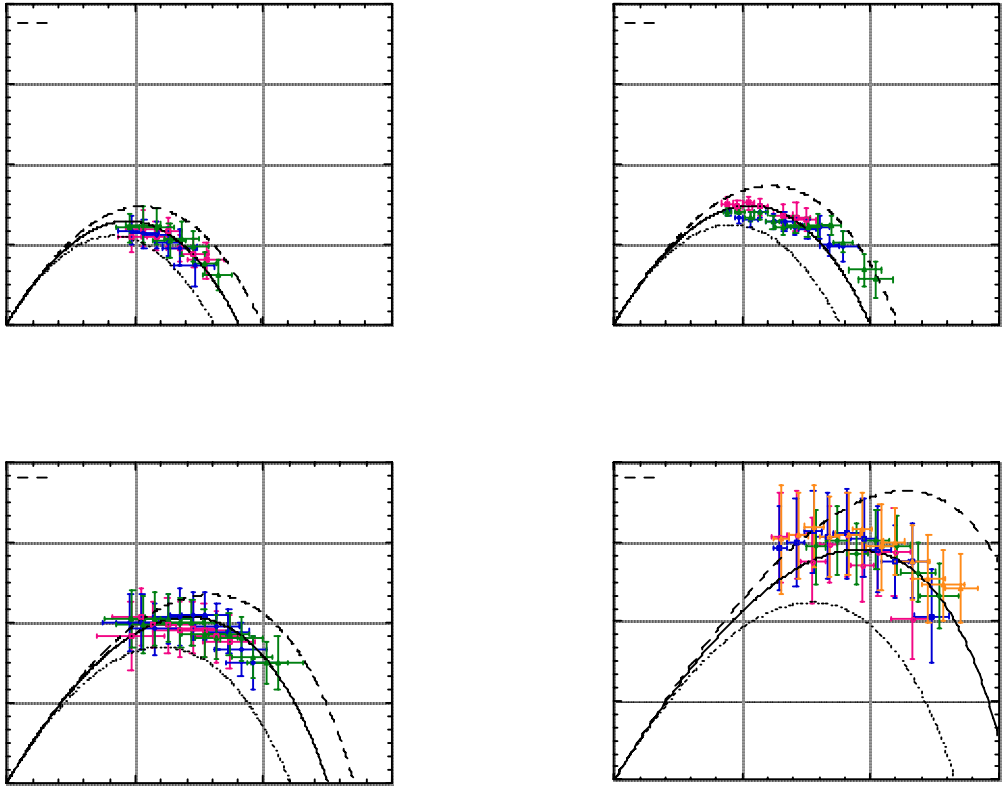

Figure 12 Measured growth rate of Darrieus-Landau instability as a function of imposed wavenumber for propane-air flames at four different equivalence ratios. The curves are calculated from (3).

The spatial growth rate was converted to a temporal growth rate from knowledge of the convection velocity of the structures, $\sigma = \sigma_x \, u_{//}$. The non-dimensional temporal growth rates were plotted as a function of the non-dimensional wavenumber in Fig. 12. The dimensional growth rate was typically 150-300 s$^{-1}$ for propane air flames. The minimum value of $f/\sigma$ was 3. Similar plots for the oxygen-enriched flames are shown in Fig.13.

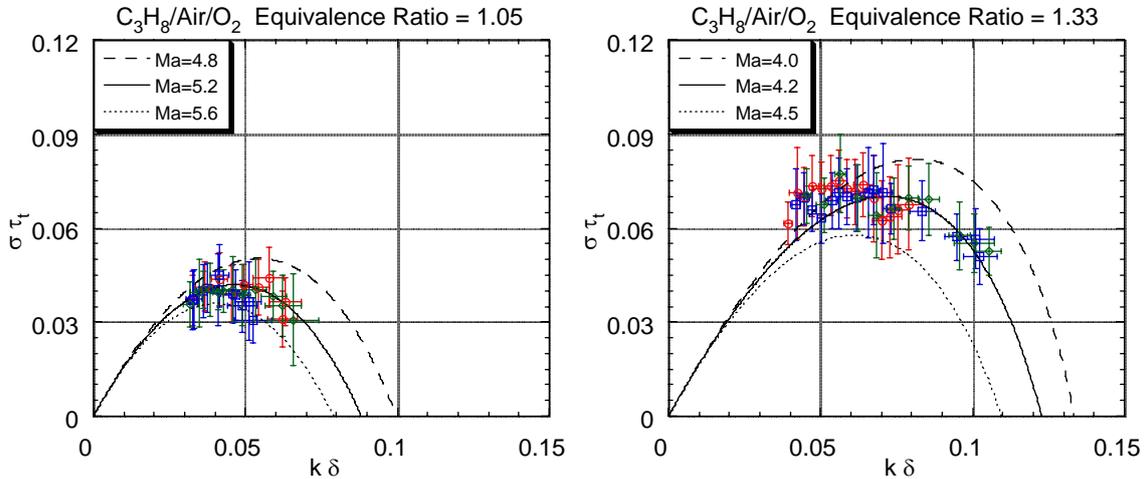

Figure 13. Measured growth rate of Darrieus-Landau instability as a function of imposed wavenumber for propane flames with 28 % oxygen at two different equivalence ratios. The curves are calculated from (3).

These experimental results are compared to the theoretical dispersion relation (3) of Clavin and Garcia [12] where the acceleration of gravity been put equal to zero. This approximation is justified because the Froude number, based on the flame speed and gravity, is always greater than 90 in these experiments. The numerical data used for the calculations is listed in table 1.

The Markstein number of each flame has been treated as an unknown parameter. For each flame we have plotted the theoretical curve for the Markstein number that agrees best with the measured results. To give an idea of the variation of the theoretical curves with this parameter, we have also plotted curves for the best values of *Ma* ± 0.3 or ± 0.2 according to sensitivity. It can be seen that the theoretical curve and the experimental measurements agree to within experimental uncertainty. It can also be seen that, for large wavenumbers, the measured growth rate of the instability decreases with increasing wavenumber, as predicted by theory including the effects of both hydrodynamics and preferential diffusion. These experimental values of the Markstein number are comparable with other values given in the literature for rich propane-air flames [14, 31].

|  | 21 % $O_2$ | | | | 28 % $O_2$ | |
| --- | --- | --- | --- | --- | --- | --- |
| $\Phi$ | 1.05 | 1.15 | 1.25 | 1.33 | 1.05 | 1.33 |
| $S_L$ m/s | 0.43 | 0.41 | 0.35 | 0.27 | 0.69 | 0.51 |
| $T_b$ K | 2067 | 2082 | 2066 | 2033 | 2334 | 2326 |
| $\delta$ μm | 47.6 | 49.6 | 57.7 | 74.4 | 28.9 | 38.1 |
| $\tau_t$ μs | 111 | 121 | 165 | 276 | 41.8 | 74.8 |
| $E$ | 7.05 | 7.11 | 7.05 | 6.94 | 7.97 | 7.94 |
| $J$ | 3.97 | 3.98 | 3.94 | 3.87 | 4.36 | 4.29 |
| $H$ | 1.108 | 1.106 | 1.089 | 1.067 | 1.198 | 1.164 |

Table 1. Numerical values used to calculate the curves in Figs.12 and 13

These experiments were limited to wave lengths less than 6 mm to avoid interaction between the two flame fronts. This constraint imposed a lower limit on the reduced wavenumber, $k\,\delta$, which was thus limited to the range 0.04–0.15. This value is smaller than unity, but not very much smaller, so the results are at the limit of the domain of validity of linearised asymptotic flame theory ($k\,\delta \ll 1$). It may also be noted that the maximum reduced growth rate, $\sigma_t$, is 0.1 for the richest propane-air flame. The results are thus also at the limit of validity low of the low frequency assumption.

**Nonlinear saturation of the instability**

In the previous sections we have been concerned with the small amplitude linear growth rate of the instability. It was seen that the growth rate is well described by linearised asymptotic laminar flame theory. In this section we will be concerned with the nonlinear effects leading to the saturation of the amplitude of the structures, see Figs 9 and 11.

The origin of the saturation mechanism is easily understood qualitatively. Consider a wrinkled flame, flat on average, propagating into a quiescent fresh mixture. At all points, the flame propagates in the direction of the local normal to the flame front. When the aspect ratio of the structures formed by the instability is not small, the local direction of propagation of the flame front is significantly different from the average plane of the flame. It is easy to see that an initially sinusoidal structure will distort. The radius of curvature of zones that are convex towards the unburned gas will increase, whereas the radius of curvature of concave zones will decrease until a cusp is formed.

A simple model equation for the evolution of the flame shape, including this geometrical nonlinearity leading to amplitude saturation, has been proposed by Sivashinsky [9, 32] in the limit of small gas expansion ($E-1 \ll 1$). This equation, known as the Michelson-Sivashinsky equation, for the time evolution of the wrinkled flame, can be written in a simplified dimensional form:

$$\frac{\partial \alpha}{\partial t} = \frac{E-1}{E} \frac{S_L}{2} \left( \frac{1}{k_n} \frac{\partial^2 \alpha}{\partial x^2} + I(\alpha, x) \right) - \frac{S_L}{2} \left( \frac{\partial \alpha}{\partial x} \right)^2, \qquad (11)$$

where $\alpha(x,t)$ is the local position of the wrinkled front with respect to some reference plane, $x$ is the transverse coordinate and $k_n$ is the neutral wavenumber for which the growth rate $\sigma(k)$ is zero. The linear operator $I(\alpha, x)$ multiplies $\alpha$ by $|k|$ in wavenumber space. It corresponds to the Darrieus-Landau instability.

$$I(e^{ikx}, x) = |k| e^{ikx} \qquad (12)$$

The second derivative, $(\partial^2 \alpha / \partial x^2)$, describes the change in local burning velocity with curvature and stretch [8, 33]. The nonlinear $x$-dependant term, $(\partial \alpha / \partial x)^2$, arises from inclination of the flame front and together with the curvature term, $(\partial^2 \alpha / \partial x^2)$, will lead to saturation. Equation (11) was obtained in the limit of Lewis number close to unity, and small gas expansion ratio, $E-1 \ll 1$. The acceleration of gravity is not considered here. In real flames the gas expansion ratio is not a small parameter, typically $E \approx 6$. Using the analytical results obtained by Boury [34, 35], Searby, Truffaut and Joulin [25] have argued that the following extension to the Michelson-Sivashinsky equation is a better model equation for the evolution of wrinkled flames when the gas expansion is not small compared to unity:

$$\frac{\partial \alpha}{\partial t} = \Omega\, S_L \left( \frac{1}{k_n} \frac{\partial^2 \alpha}{\partial x^2} + I(\alpha, x) \right) - a \frac{S_L}{2} \left( \frac{\partial \alpha}{\partial x} \right)^2$$

$$\Omega = \frac{E}{E+1} \sqrt{\frac{E^2 + E - 1}{E}} - 1, \quad a = \frac{2E}{E-1}$$

(13)

here $\Omega$ is the constant of proportionality found by Landau, see equ (1). Equation (13) reduces to (12) in the limit $E-1 \ll 1$. Searby, Truffaut and Joulin [25] have further adapted equ (13) to the case of an anchored flame with nonzero tangential flow velocity. They propose the following model equation to describe the growth of cellular structures with arbitrary gas expansion and convected by a tangential velocity $u_{//}$:

$$\frac{\partial \alpha}{\partial t} + u_{//} \frac{\partial \alpha}{\partial x} = \Omega\, S_L \left( \frac{1}{k_n} \frac{\partial^2 \alpha}{\partial x^2} + I(\alpha, x) \right) - a \frac{S_L}{2} \left( \frac{\partial \alpha}{\partial x} \right)^2 + \frac{S_L}{2}(a-1) \left\langle \left( \frac{\partial \alpha}{\partial x} \right)^2 - V\left( \frac{x}{u_{//}} \right) \right\rangle$$

(14)

The time averaged values of the increase in arc length, $(1/2)\left\langle (\partial \alpha / \partial x)^2 \right\rangle$, and the time averaged normal velocity of the fresh gases are considered to depend only on $x / u_{//}$. Since neither term appears in the linear analysis, they are at least quadratic in the amplitude of wrinkling.

Using a pole decomposition scheme and adapting the results of references [36, 37], it can be shown that (14) has an exact solution:

$$\alpha(\chi, \tau) = -A(\tau) - \frac{2\Omega}{a\, k_n} \ln\left[1 - \frac{\cos(K(\chi - \chi_0))}{\cosh(K\, B)}\right], \tag{15}$$

where $\Omega$ and $a$ are defined above, $\chi = x - u_{//}\, t$ and $\tau = x\, u_{//}$ are the Lagrangian distance and time respectively, $\chi_0$ and $K$ are constants, and the functions $A$ and $B$ are solutions to the coupled ordinary differential equations:

$$\begin{aligned}\frac{dA}{dt} &= S_L\left[\left(\frac{2\Omega\, K}{a\, k_n}\right)^2 \frac{\exp(-2K\, B)}{1 - \exp(-2K\, B)}\right] \\ \frac{dB}{dt} &= S_L\, \Omega\left[\frac{K}{k_n}\coth(K\, B) - 1\right]\end{aligned} \tag{16}$$

The function $B \to \infty$ for $\tau \to -\infty$. Admittedly (15-16) is not the most general known exact solution to (14) (see Refs. [36, 37]), but it is the only relevant one if $k_n/3 < K \leq k_n$ when starting from an infinitesimal $2\pi/K$-periodic pattern [38]. At fixed $x$, the above solution for $\alpha(\chi, \tau)$ oscillates in time with a pulsation $\omega = u_{//} K$. This yields $K$ if $u_{//}$ and $\omega$ are known. The local maxima and minima of $\alpha(\chi, \tau)$ occur when $\cos(K(\chi - \chi_o)) = \pm 1$. The peak to peak amplitude of wrinkling, $\alpha_{p-p}$, is thus easily found:

$$\alpha_{p-p}(x) = \frac{2\Omega}{a\, k_n} \ln\left\{\frac{\cosh[K\, B(x/u_{//})] + 1}{\cosh[K\, B(x/u_{//})] - 1}\right\} \tag{17}$$

This result can easily be compared with the experimental observations in Fig.11. The unknown quantities in (17) and (16) are $a(E)$, $\Omega(E)$ and $k_n$. The parameters $a(E)$ and $\Omega(E)$ were evaluated using the expansion coefficients in table 1. The parameter $k_n$ was obtained from the experimental measurement of the dispersion relation, Fig.13, by adjusting a parabola $\sigma \tau_t = (\Omega k / k_n)(1 - k / k_n)$ through the measured points. The mean tangential flow velocity, $u_{//}$, was 8.59 m/s.

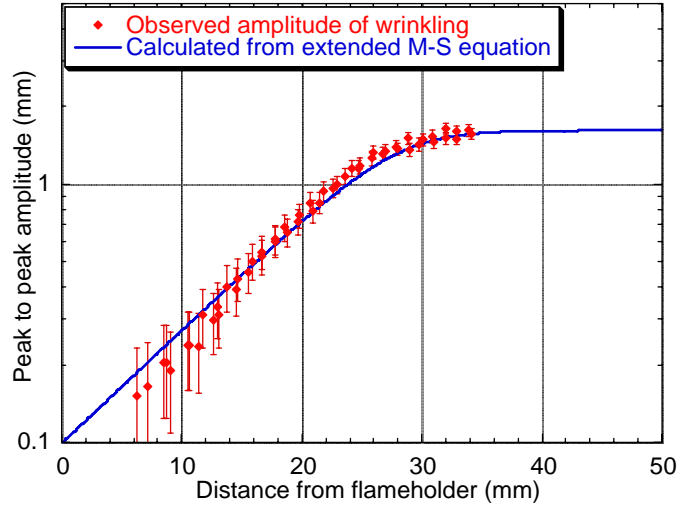

Figure 14. Comparison of prediction of eq.(17) with the data from Fig 11. $\Omega(E) = 1.79$, $a(E) = 1.78$, $k / k_n = 0.38$.

The result of the calculation is shown in Fig.14, along with the experimental points taken from Fig.11. The only adjustable parameter in this plot is the initial amplitude of wrinkling. All the other parameters are predetermined. It is not really surprising that the exponential part of the growth is well represented, this just means that, to order $k^2$, the Michelson-Sivashinsky equation (14) has the same dispersion relation as the Clavin-Garcia equation (3-4). It is however remarkable that the Michelson-Sivashinsky equation also correctly predicts both the saturation amplitude of the structures and the width of the cross-over region between exponential growth and constant saturated amplitude.

The complete non-linear flame profile is given by (15-16). We have used this equation to calculate the curve shown in Fig.16. We have used the same input data as in Fig.15. Extra experimental parameters are the initial flame position and the initial phase of wrinkling. The excitation frequency is 2100 Hz. As expected from Fig.15, equation (15-16) gives a good overall prediction of the growth of the instability. It also gives a good representation of the form of the wrinkling, evolving from a sinusoidal to a cusped shape. The precision of the measurements of the flame profile from Fig.9 is insufficient to make any direct comparison concerning the radius of curvature of the cusps.

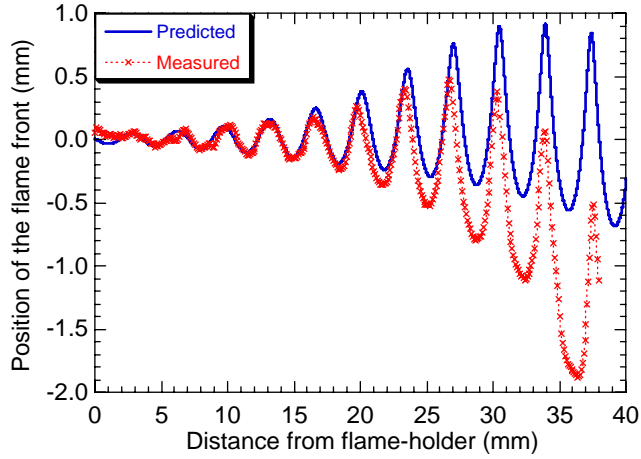

Figure 15. Comparison of prediction of (15) with the flame profile of Fig.10. Input data as in Fig.15.

However it is clear from the comparison of the observed and calculated flame profiles in Fig15, that the flame front accelerates towards the burnt gas as the wrinkling increases, and the amplitude of this acceleration is strongly underestimated by (15), in which the increase in average flame speed arises solely from the increase in the arclength of the front, given by:

$$\int_0^x \left( \sqrt{1+\left(\frac{\partial \alpha}{\partial x}\right)^2} - 1 \right) dx \approx \frac{1}{2} \int_0^x \left(\frac{\partial \alpha}{\partial x}\right)^2 dx \qquad (18)$$

The origin of this strong increase in apparent overall flame propagation speed is the production of a large-scale flow circulation induced by the increased average propagation velocity along the flame. This $x$-dependant increase in the average flame speed creates an $x$-dependant increase in the pressure jump across the flame brush, which in turn bends the overall hydrodynamic flow.

We may finally remark that this transverse circulation, will make the other, unexcited, side of our two-dimensional Bunsen flame acquire a time-averaged deflection from its straight unperturbed shape. This is indeed the case, as can be seen in Fig.5. The unexcited side of the flame is curved away from the unburned gas, and this curvature increases towards the flame tip. Accordingly, measuring the instantaneous distance between the two flame fronts, then subtracting the unperturbed value should give access to a comparison of theory and experiments that is unaffected by the global transverse circulation. This has been done in Fig.17.

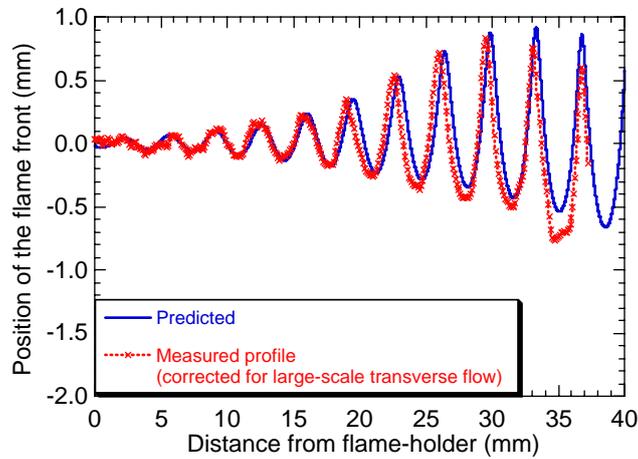

Figure17. Comparison of the predictions of (16) with the flame profile of Fig.5, after correction for the large-scale transverse flow.

This treatment noticeably improves the agreement compared to Fig.16. Experimentally the last cell is still more developed than predicted. However, considering the small distance between the two opposed flame fronts at this location, this difference can be attributed to a small-scale hydrodynamic interaction between the two flame fronts. In Fig.5, the development of a cell at the far end of the unexcited flame front is indeed a sign of such an interaction.

## Conclusion

We have recalled the mechanisms controlling the stability of planar laminar premixed flames, along with a brief historical survey of the theoretical approaches used to model them analytically. Despite the fact that the intrinsic instability of planar flames has been recognised for nearly a century and that the first attempts at an analytical description were performed over fifty years ago, experimental validation of the predicted growth rates of unstable structures has been obtained only recently. The main reason for this lies in the experimental difficulty of preparing and controlling an initially planar premixed flame front in a regime where the planar front is unstable.

We have described two novel experiments in which the growth rate of cellular structures on planar flames has been measured directly. In the first experiment an unstable laminar flame front is maintained initially planar by the action of an imposed acoustic field. The temporal growth rate of 2-D wrinkling is then observed after removal of the stabilising acoustic field. In the second experiment the spatio-temporal growth of wrinkling is observed on an inclined anchored flame. The residual turbulence of the gas flow is sufficiently low that turbulence induced perturbations at the base of the flame are not amplified to a perceptible amplitude before reaching the tip of the flame. Perturbations of controlled wavelength and amplitude are then created by an electrostatic deflection system. This technique has permitted the exploration of a significant portion of the dispersion curve. The spatial growth rate of the wrinkling was converted to a temporal growth rate using a Lagrangian time. These experimental observations have confirmed the validity of the analytical theory.

In the final section of this paper we have investigated the non-linear saturation of the instability. Our experimental observations have been compared to the predictions of a Michelson-Sivashinsky equation, extended to the case of strong gas expansion and adapted to include tangential convective flow. It was seen that the extended equation correctly describes both the amplitude of structures at saturation and also the transition region from exponential growth to saturation. The experimental observations show a strong apparent overall increase in velocity of the wrinkled flame front. The origin of this effect is attributed to the generation of a large-scale hydrodynamic flow, induced by the x-dependence of the pressure jump across the wrinkled flame. This effect is absent from the Michelson-Sivashinsky description.

## Acknowledgements

The author is indebted to G. Joulin for his helpful discussions and suggestions. He also acknowledges the substantial contributions of C. Clanet and J-M Truffaut. Without the invaluable technical assistance J. Minelli and F. Abetino much of this work could not have been completed. Part of this work was carried out with the financial support of Air Liquide Welding.

## References

[1]  Darrieus, G., (1938), Propagation d'un front de flamme. Unpublished work presented at La Technique Moderne, and at Le Congrés de Mécanique Appliqueée (1945).

[2] Landau, L., (1944), On the theory of slow combustion. Acta Phys.-Chim. URSS, 19, 77-85.

[3]  Zeldovich, Y.B., (1944), The Theory of Combustion and Detonation of Gases. Akademiia Nauk SSSR, Moscow.

[4] Zeldovich, Y.B. & Barenblatt, G.I., (1959), Theory of flame propagation. Combust. Flame, 3, 61-74.

[5] Barenblatt, G.I., Zeldovich, Y.B. & Istratov, A.G., (1962), On the Diffusional Thermal Stability of Laminar Flames. Prikl. C. Mekh. Fiz., 4, 21.

[6]  Sivashinsky, G.I., (1977), Diffusional-thermal theory of cellular flames. Combust. Sci. Technol., 15, 137-146.

[7]  Joulin, G. & Clavin, P., (1979), Linear stability analysis of nonadiabatic flames : Diffusional-thermal model. Combust. Flame, 35, 139-153.

[8]  Markstein, G.H., (1964), Nonsteady flame propagation. Pergamon, New York.

[9]  Sivashinsky, G.I., (1977), Nonlinear analysis of hydrodynamic instability in laminar flames-I. Derivation of basic equations. Acta Astronaut., 4, 1177-1206.

[10] Clavin, P. & Williams, F.A., (1982), Effects of molecular diffusion and of thermal expansion on the structure and dynamics of premixed flames in turbulent flows of large scale and low intensity. J. Fluid Mech., 116, 251-282.

[11] Pelcé, P. & Clavin, P., (1982), Influence of hydrodynamics and diffusion upon the stability limits of laminar premixed flames. J. Fluid Mech., 124, 219-237.

[12]  Clavin, P. & Garcia, P., (1983), The influence of the temperature dependence of diffusivities on the dynamics of flame fronts. J. Méc. Théor. Appl., 2, 245-263.

[13]  Searby, G. & Quinard, J., (1990), Direct and indirect measurements of Markstein numbers of premixed flames. Combust. Flame, 82, 298-311.

[14]  Tseng, L.K., Ismail, M.A. & Faeth, G.M., (1993), Laminar burning velocities and Markstein numbers of hydrocarbon/air flames. Combust. Flame, 95, 410-426.

[15]  Deshaies, B. & Cambray, P., (1990), The velocity of a premixed flame as a function of flame stretch : An experimental study. Combust. Flame, 82, 361-375.

[16]  Bradley, D., Hicks, R.A., Lawes, M., Sheppard, C.G.W. & Woolley, R., (1998), The measurement of laminar burning velocities and Markstein numbers for iso-octane and iso-octane--n-heptane-- air mixtures at elevated temperatures and pressures in an explosion bomb. Combust. Flame, 115, 126-144.

[17]  Davis, S.G., Quinard, J. & Searby, G., (2001), A numerical investigation of stretch effects in counterflow premixed laminar flames. Combustion Theory and Modelling, 5, 353-362.

[18]  Davis, S.G., Quinard, J. & Searby, G., (2002), Markstein numbers in counterflow, methane- and propane-air flames: a computational study. Combust. Flame, 130, 123-136.

[19]  Kee, R.J., Rupley, F.M. & Miller, J.A., (1990), The Chemkin thermodynamic data base. SAND87-8215B, Report, Sandia National Laboratories.

[20]  Clanet, C. & Searby, G., (1998), First experimental study of the Darrieus-Landau instability. Phys. Rev. Lett., 27, 3867-3870.

[21]  Searby, G. & Rochwerger, D., (1991), A parametric acoustic instability in premixed flames. J. Fluid Mech., 231, 529-543.

[22]  McLachlan, N.W., (1951), Theory and Application of Mathieu Functions. Clarendon, Oxford.

[23]  Yamaoka, I. & Tsuji, H., (1984), Determination of burning velocity using counterflow flames. Proc. Combust. Inst., 20, 1883-1892.

[24]  Truffaut, J.M. & Searby, G., (1999), Experimental study of the Darrieus-Landau instability on an inverted-`V' flame and measurement of the Markstein number. Combust. Sci. Technol., 149, 35-52.

[25]  Searby, G., Truffaut, J.M. & Joulin, G., (2001), Comparison of experiments and a non-linear model for spatially developing flame instability. Phys. Fluids, 13, 3270-3276.

[26]  Garcia, P. & Borghi, R., (1986), Etude de la stabilité de flammes prémélangées obliques. Journal of Theoretical and Applied Mechanics, Special Issue, 157-172.

[27]  Polanyi, M.L. & Markstein, G.H., (1947), Phenomena in electrically and acoustically disturbed Bunsen burner flames. 5, Project SQUID technical report, Cornell Aeronautical Laboratory.

[28]  Payne, K.G. & Weinberg, F.G., (1958), A preliminary investigation of field-induced ion movement in flame gases and its applications. Proc. R. Soc. London, A250, 316-336.

[29]  Bradley, D., (1986), The effects of electric fields on combustion processes. Advanced Combustion Methods, pp. 331-394. Academic Press,

[30]  Clavin, P., (1985), Dynamic behaviour of premixed flame fronts in laminar and turbulent flows. Prog. Energy Comb. Sci., 11, 1-59.

[31]  Kwon, S., Tseng, L.K. & Faeth, G.M., (1992), Laminar burning velocities and transition to unstable flames in H2/O2/N2 and C3H8/O2/N2 mixtures. Combust. Flame, 90, 230-246.

[32]  Michelson, D.M. & Sivashinsky, G.I., (1977), Nonlinear analysis of hydrodynamic instability in laminar flames-II. Numerical experiments. Acta Astronaut., 4, 1207-1221.

[33] Clavin, P. & Joulin, G., (1983), Premixed flames in large scales and high intensity turbulent flow. J. Phys. Lett. (Paris), 44, L-112.

[34] d'Angelo, Y., Joulin, G. & Boury, G., (2000), On model evolution equations for the whole surface of 3-D expanding wrinkled premixed flames. Combustion Theory and Modelling, 4, 317-338.

[35] Boury, G., (2003), Études théoriques et numériques de fronts de flammes plissées : Dynamiques non-linéaires libres ou bruitées. PhD, Université de Poitiers.

[36] Lee, Y.C. & Chen, H.H., (1982), Nonlinear dynamical models of plasma turbulence. Physica Scripta (Sweden), 2, 41.

[37] Thual, O., Frisch, U. & Henon, M., (1985), Application of pole decomposition to an equation governing the dynamics of wrinkled flame fronts. Journal de Physique (France), 46, 1485-1494.

[38] Joulin, G. & Cambray, P., (1992), On a tentative approximate evolution equation for markedly wrinkled premixed flames. Combust. Sci. Technol., 81, 243-256.